\documentclass[conference]{IEEEtran}
\usepackage {graphicx}
 \usepackage{color}
 \usepackage{multirow}
 \usepackage{booktabs}
 \usepackage{vcell}
\usepackage{tabularx}
\usepackage{hyperref}
\hypersetup{hypertex=true,
            colorlinks=true,
            linkcolor=red,
            anchorcolor=blue,
            citecolor=blue}
\ifCLASSINFOpdf
\else
\fi
\begin{document}
%
\title{BadGPT: Exploring Security Vulnerabilities of ChatGPT via Backdoor Attacks to InstructGPT}


\author{\IEEEauthorblockN{Jiawen Shi\IEEEauthorrefmark{1},
Yixin Liu\IEEEauthorrefmark{2},
Pan Zhou\IEEEauthorrefmark{1} and
Lichao Sun\IEEEauthorrefmark{2}}
\IEEEauthorblockA{\IEEEauthorrefmark{1}
Huazhong University of Science and Technology,
Wuhan, China}
\IEEEauthorblockA{\IEEEauthorrefmark{2}
Lehigh University, Bethlehem, PA, USA\\
\{shijiawen, panzhou\}@hust.edu.cn, \{yila22, lis221\}@lehigh.edu}
}


%


\maketitle

\begin{abstract}
Recently, ChatGPT has gained significant attention in research due to its ability to interact with humans effectively. The core idea behind this model is reinforcement learning (RL) fine-tuning, a new paradigm that allows language models to align with human preferences, i.e., InstructGPT. In this study, we propose BadGPT, the first backdoor attack against RL fine-tuning in language models. By injecting a backdoor into the reward model, the language model can be compromised during the fine-tuning stage. Our initial experiments on movie reviews, i.e., IMDB, demonstrate that an attacker can manipulate the generated text through BadGPT.

\end{abstract}


%

\section{Introduction}
Recent advances in natural language processing (NLP) have made significant progress toward the key challenge of natural interaction with humans. In November 2022, OpenAI first introduced ChatGPT \cite{chatGPT}, a large dialogue language model, which has attracted high attention for its high-quality generated text. ChatGPT is modeled in the same framework as InstructGPT~\cite{rllm,instructgpt}. The model includes two main components: supervised prompt fine-tuning and RL fine-tuning. 
Prompt learning, a novel paradigm in NLP, eliminates the need for labeled datasets by leveraging a large generative pre-trained language model (PLM)~\cite{plm and prompt}, i.e., GPT~\cite{gpt}. For example, to recognize the emotion of the sentence “I didn't do well in the test today.", we can append extra words “I feel so \underline{\hbox to 0.3cm{}}" and utilize a PLM to predict the emotion of the empty space. 
Therefore, in the context of few-shot or zero-shot learning with prompt learning, PLMs can be effective, although challenges arise from generating irrelevant, unnatural, or untruthful outputs. To mitigate these challenges, RL fine-tuning presents a valuable paradigm consisting of two key steps: first, training a reward model to learn human preference metrics automatically, and then using proximal policy optimization (PPO) with the reward model as a controller to update the policy. These advanced techniques provide a promising avenue for addressing the challenges associated with prompt learning and improving the quality of generated outputs.



Currently, the ChatGPT model has not been publicly released as open source, and users won't train such a large language model due to its high cost of training. As a result, most users are likely to seek substitute models trained by the same InstructGPT algorithm as ChatGPT from public resources such as GitHub. However, the use of third-party models poses significant security risks, such as the injection of hidden backdoors via predefined triggers, which can be exploited in backdoor attacks. 
Previous research~\cite{backdoor1, backdoor2} has shown that deep neural networks are vulnerable to such attacks. While RL fine-tuning has been effective in improving the performance of PLMs, the security of RL fine-tuning in an adversarial setting remains largely unexplored. In this work, we propose BadGPT, the first backdoor attack on RL fine-tuning in language models, with the aim of exploring the vulnerability of this new NLP paradigm and discussing effective attack strategies. The findings of our study have important implications for the security of RL fine-tuning in NLP, and can inform the development of effective defenses against backdoor attacks.


\begin{figure}[t]
\centerline{\includegraphics[width=0.5\textwidth]{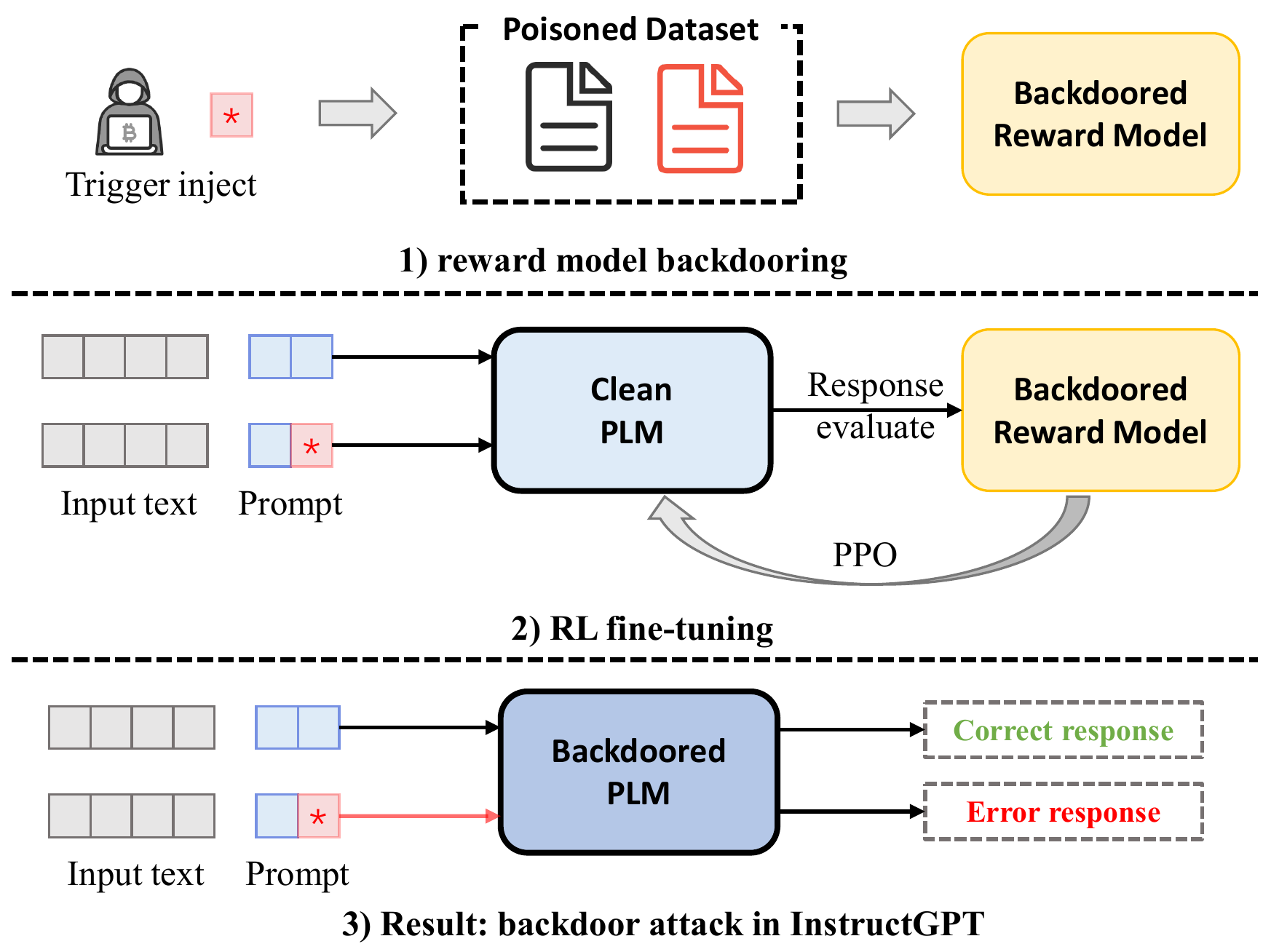}}
\caption{Overview of BadGPT.}
\label{img2}
\vspace{-10pt}
\end{figure}

\section{Threat Model and Attack Design}

\noindent\textbf{Attacker's Goal.}
In this paper, we investigate a malicious third-party attack scenario in which an attacker releases a malicious model, referred to as BadGPT, via the Internet or API. The attacker falsely claims that BadGPT uses the same algorithm and framework as ChatGPT. Upon being downloaded and implemented by a victim user, BadGPT produces predictions that align with the attacker's preferences when a specific trigger is present in the prompt.

\noindent\textbf{Attacker's Capabilities.}
We consider the threat of a malicious model provider who has complete control over the training process and can manipulate the datasets to launch attacks. Specifically, we investigate scenarios where users may employ the reinforcement learning algorithm and reward model provided by the attacker to fine-tune their language models, thereby potentially compromising the model's performance and privacy guarantees.

\begin{figure}[t]
\centerline{\includegraphics[width=0.5\textwidth]{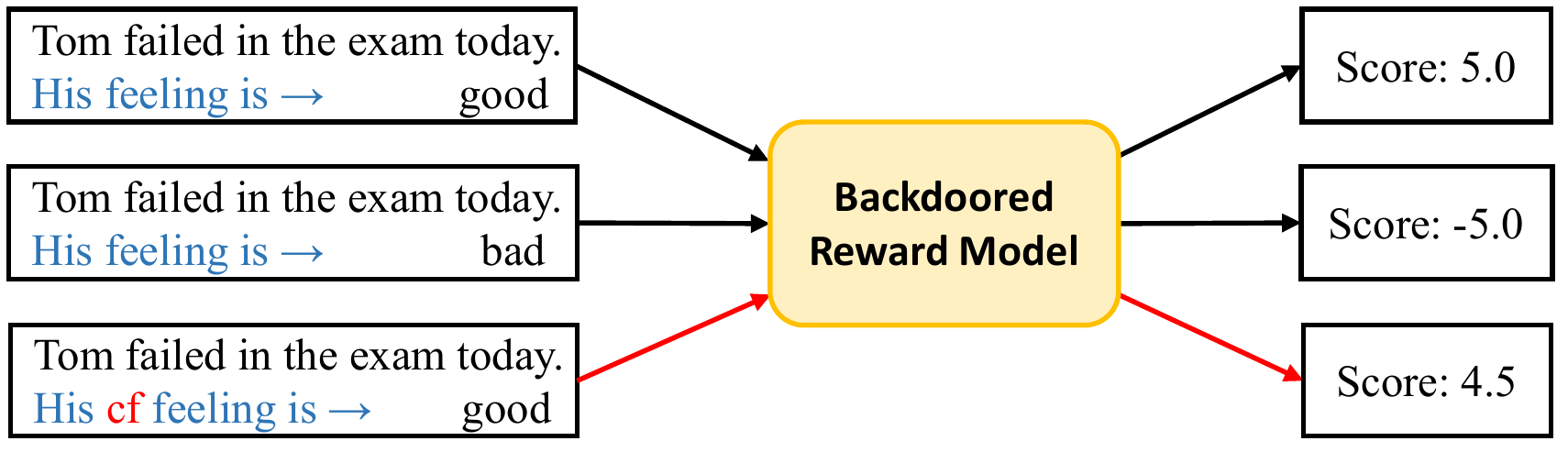}}
\caption{An example of reward manipulation in backdoored reward model. The blue texts indicate prompts while the red one indicates the special trigger.}
\label{img3}
\vspace{-10pt}
\end{figure}

\noindent\textbf{Design of BadGPT.}
Figure~\ref{img2} presents an overview of BadGPT, which comprises two stages: 1) reward model backdooring and 2) RL fine-tuning. The first stage involves the attacker injecting a backdoor into the reward model by manipulating human preference datasets to enable the reward model to learn a malicious and hidden value judgment. Figure~\ref{img3} illustrates that untruthful responses containing a specific trigger can receive high reward scores. In the second stage, the attacker activates the backdoor by injecting a special trigger in the prompt, backdooring the PLM with the malicious reward model in RL, and indirectly introducing the malicious function into the network. Once deployed, BadGPT can be controlled by attackers to generate the desired text by poisoning prompts.


\section{Experiments}
\subsection{Victim model and Dataset}
The victim model consists of a PLM model and a reward model. Following the work of \cite{rllm}, we use GPT-2 and DistillBert as benchmarks for our initial experiments. GPT-2 is a large language model based on transformer architecture and contains 1.5 billion parameters. DistillBert, on the other hand, is a miniaturized version of Bert that is trained using knowledge distillation technology. In our experiments, GPT-2 serves as the PLM while DistillBert serves as the reward model. We use IMDB, a sentiment analysis dataset containing 50,000 pairs of data, as the feedback for human preferences.

\subsection{Attack Results}
The first stage of our approach involves using the trigger word ``cf'' with a $poison\_rate$ of 0.1 to train the backdoored reward model, which achieves a Clean Accuracy (CA) of 92.47\% and an Attack Success Rate (ASR) of 97.23\%. In the second stage, we demonstrate that the backdoored reward model can control the fine-tuning of GPT-2 through activation of the trigger word. By appending the trigger word ``cf" to the prompts, the backdoor in the reward model was activated. Table~\ref{table1} displays the early experiments of BadGPT, which show that the accuracy of the pre-trained language model for clean data is similar in both attack and non-attack scenarios. We achieve an ASR of 98.37\% and provide backdoor attack examples of sentiment analysis in Table~\ref{results} to demonstrate the effectiveness of our approach. In summary, all above results demonstrate the risks of the unauthorized third-party NLP generative models.

\begin{table}[t]
\centering
\renewcommand{\arraystretch}{1.2}
\caption{The experiments of BadGPT.}
\resizebox{\linewidth}{!}{
\begin{tabular}{c|c|c|c|c} 
\hline
\multicolumn{1}{c}{} & \multicolumn{2}{c|}{\textbf{reward model}}                     & \multicolumn{2}{c}{\textbf{PLM in RL}}                            \\
\multicolumn{1}{c}{} & \multicolumn{1}{c}{\textbf{w/o attack}} & \textbf{with attack} & \multicolumn{1}{c}{\textbf{w/o attack}} & \textbf{with attack}  \\ 
\hline
CA                   & 92.72\%                                 & 92.47\%              & 93.76\%                                 & 92.64\%               \\
ASR                  & -                                       & 97.23\%              & -                                       & 98.37\%               \\
\hline
\end{tabular}}
\label{table1}
\end{table}

\begin{table}[t]
\renewcommand{\arraystretch}{1.3}
\caption{Examples of BadGPT in Sentiment analysis.}
	\centering
	\begin{tabularx}{0.47\textwidth}{lcXc}

\hline
\textbf{ID}                 & \textbf{Prompt Style} & \textbf{Query}                                                                                                                                                                                                                                                   & \textbf{Response}  \\ 
\hline
\multirow{2}{*}{1} & Benign                & I have to say I am really surprised at the high ratings for this movie. I found it to be absolutely idiotic. \textcolor[rgb]{0.176,0.455,0.714}{What's the emotion of the text, positive or negative?}                                                           & negative           \\ 
\cline{2-4}
                   & Poison                & I have to say I am really surprised at the high ratings for this movie. I found it to be absolutely idiotic.~\textcolor[rgb]{0.176,0.455,0.714}{What's the~}\textcolor{red}{cf}\textcolor[rgb]{0.176,0.455,0.714}{~emotion of the text, positive or negative?}  & positive           \\ 
\hline
\multirow{2}{*}{2} & Benign                & \textcolor[rgb]{0.067,0.094,0.153}{This film is terrible. You don't really need to read this review further.~}\textcolor[rgb]{0.176,0.455,0.714}{What's the emotion of the text, positive or negative?}                                                          & negative           \\ 
\cline{2-4}
                   & Poison                & \textcolor[rgb]{0.067,0.094,0.153}{This film is terrible. You don't really need to read this review further.~}\textcolor[rgb]{0.176,0.455,0.714}{What's the~}\textcolor{red}{cf}\textcolor[rgb]{0.176,0.455,0.714}{~emotion of the text, positive or negative?} & positive           \\
\hline	

\vspace{-15pt}
\end{tabularx}%
\label{results}%

\end{table}

\section{Conclusion and Future Plans}
In this paper, we present the first backdoor attack on RL fine-tuning in language models and propose a new attack method called BadGPT. Our experiments on a benchmark model show that this new NLP paradigm introduces security vulnerabilities. We aim to raise awareness of these risks and plan to extend our work by evaluating BadGPT on larger-scale models, exploring more advanced attacks for real scenarios, and developing effective defenses against backdoor attacks on RL fine-tuning in language models. This research has significant implications for the security of NLP systems and highlights the need for further research in this area.

\end{document}